\documentstyle[prl,preprint,aps,epsfig,amssymb,amsfonts]{revtex}

\begin{document}

\draft

\title{\boldmath
First Measurement of the Ratio  
$B(t\rightarrow Wb)/B(t\rightarrow Wq)$ and Associated Limit on
the CKM Element $|V_{tb}|$ }

\maketitle

\font\eightit=cmti8
\def\r#1{\ignorespaces $^{#1}$}
\hfilneg
\begin{sloppypar}
\noindent
T.~Affolder,\r {23} H.~Akimoto,\r {45}
A.~Akopian,\r {38} M.~G.~Albrow,\r {11} P.~Amaral,\r 8 
S.~R.~Amendolia,\r {34} 
D.~Amidei,\r {26} K.~Anikeev,\r {24} J.~Antos,\r 1 
G.~Apollinari,\r {11}
% ...
\end{sloppypar}
\vskip .026in
\begin{center}
(CDF Collaboration)
\end{center}

\vskip .026in
\begin{center}
\r 1  {\eightit Institute of Physics, Academia Sinica, Taipei, Taiwan 11529, 
Republic of China} \\
\r 2  {\eightit Argonne National Laboratory, Argonne, Illinois 60439} \\
\r 3  {\eightit Istituto Nazionale di Fisica Nucleare, University of Bologna,
I-40127 Bologna, Italy} \\
\r 4  {\eightit Brandeis University, Waltham, Massachusetts 02254} \\
\r 5  {\eightit University of California at Davis, Davis, California  95616} \\
\r 6  {\eightit University of California at Los Angeles, Los 
Angeles, California  90024} \\  
\r 7  {\eightit Instituto de Fisica de Cantabria, 
CSIC-University of Cantabria, 
39005 Santander, Spain} \\
\r 8  {\eightit Enrico Fermi Institute, University of Chicago, Chicago, 
Illinois 60637} \\
\r 9  {\eightit Joint Institute for Nuclear Research, RU-141980 Dubna, Russia}
\\
\r {10} {\eightit Duke University, Durham, North Carolina  27708} \\
\r {11} {\eightit Fermi National Accelerator Laboratory, Batavia, Illinois 
60510} \\
\r {12} {\eightit University of Florida, Gainesville, Florida  32611} \\
\r {13} {\eightit Laboratori Nazionali di Frascati, 
Istituto Nazionale di Fisica
               Nucleare, I-00044 Frascati, Italy} \\
\r {14} {\eightit University of Geneva, CH-1211 Geneva 4, Switzerland} \\
\r {15} {\eightit Glasgow University, Glasgow G12 8QQ, United Kingdom}\\
\r {16} {\eightit Harvard University, Cambridge, Massachusetts 02138} \\
\r {17} {\eightit Hiroshima University, Higashi-Hiroshima 724, Japan} \\
\r {18} {\eightit University of Illinois, Urbana, Illinois 61801} \\
\r {19} {\eightit The Johns Hopkins University, Baltimore, Maryland 21218} \\
\r {20} {\eightit Institut f\"{u}r Experimentelle Kernphysik, 
Universit\"{a}t Karlsruhe, 76128 Karlsruhe, Germany} \\
\r {21} {\eightit Center for High Energy Physics: Kyungpook National
University, Taegu 702-701; Seoul National University, Seoul 151-742; and
SungKyunKwan University, Suwon 440-746; Korea} \\
\r {22} {\eightit High Energy Accelerator Research Organization (KEK), 
Tsukuba, 
Ibaraki 305, Japan} \\
\r {23} {\eightit Ernest Orlando Lawrence Berkeley National Laboratory, 
Berkeley, California 94720} \\
\r {24} {\eightit Massachusetts Institute of Technology, Cambridge,
Massachusetts  02139} \\   
\r {25} {\eightit Institute of Particle Physics: McGill University, Montreal 
H3A 2T8; and University of Toronto, Toronto M5S 1A7; Canada} \\
\r {26} {\eightit University of Michigan, Ann Arbor, Michigan 48109} \\
\r {27} {\eightit Michigan State University, East Lansing, Michigan  48824} \\
\r {28} {\eightit University of New Mexico, Albuquerque, New Mexico 87131} \\
\r {29} {\eightit The Ohio State University, Columbus, Ohio  43210} \\
\r {30} {\eightit Osaka City University, Osaka 588, Japan} \\
\r {31} {\eightit University of Oxford, Oxford OX1 3RH, United Kingdom} \\
\r {32} {\eightit Universita di Padova, Istituto Nazionale di Fisica 
          Nucleare, Sezione di Padova, I-35131 Padova, Italy} \\
\r {33} {\eightit University of Pennsylvania, Philadelphia, 
        Pennsylvania 19104} \\   
\r {34} {\eightit Istituto Nazionale di Fisica Nucleare, University and Scuola
               Normale Superiore of Pisa, I-56100 Pisa, Italy} \\
\r {35} {\eightit University of Pittsburgh, Pittsburgh, Pennsylvania 15260} \\
\r {36} {\eightit Purdue University, West Lafayette, Indiana 47907} \\
\r {37} {\eightit University of Rochester, Rochester, New York 14627} \\
\r {38} {\eightit Rockefeller University, New York, New York 10021} \\
\r {39} {\eightit Rutgers University, Piscataway, New Jersey 08855} \\
\r {40} {\eightit Texas A\&M University, College Station, Texas 77843} \\
\r {41} {\eightit Texas Tech University, Lubbock, Texas 79409} \\
\r {42} {\eightit Istituto Nazionale di Fisica Nucleare, University of Trieste/
Udine, Italy} \\
\r {43} {\eightit University of Tsukuba, Tsukuba, Ibaraki 305, Japan} \\
\r {44} {\eightit Tufts University, Medford, Massachusetts 02155} \\
\r {45} {\eightit Waseda University, Tokyo 169, Japan} \\
\r {46} {\eightit University of Wisconsin, Madison, Wisconsin 53706} \\
\r {47} {\eightit Yale University, New Haven, Connecticut 06520} \\
\r {(\ast)} {\eightit Now at Carnegie Mellon University, Pittsburgh,
Pennsylvania  15213}
\end{center}

\begin{abstract}
We present the first measurement of the ratio of
branching fractions $R \equiv {B(t\rightarrow Wb)}/{B(t\rightarrow Wq)}$
from $p \bar{p}$ collisions at $\sqrt{s}=1.8$ TeV. The data set corresponds
to 109 pb$^{-1}$ of data recorded by the Collider Detector at Fermilab 
during the 1992-95 Tevatron run. We measure 
$R ={0.94}^{+0.31}_{-0.24}$(stat+syst) or  
$R > 0.61$ $(0.56)$ at 90 (95)\% CL,
in agreement 
with the standard model predictions. 
This measurement yields a limit of 
the Cabibbo-Kobayashi-Maskawa quark mixing matrix element $|V_{tb}|$ under
the assumption of three generation unitarity.
\end{abstract}


The Cabibbo-Kobayashi-Maskawa matrix~\cite{ckm} is a fundamental component 
of the standard model of electroweak interactions. However, the 
matrix elements must be determined experimentally since the model does not
constrain their values.
Some of the matrix elements have been determined by studying 
the weak decay of 
quarks or by deep inelastic neutrino scattering experiments. 
Until now no direct information has been available 
for the elements of the top 
sector. The matrix elements $|V_{td}|$ and $|V_{ts}|$ have 
been indirectly estimated 
and $|V_{tb}|$ has been deduced by a global fit, 
with the additional assumptions of 
having only three generations and unitarity. 
With this procedure the indirect 
allowed range for  $|V_{tb}|$ is $0.9989 \div 0.9993$ 
(at 90\% CL)~\cite{pdg}. 
However, without the assumption of
three quark generations, the constraint is far less stringent:
$|V_{tb}| = 0. \div 0.9993$ (at 90\% CL)~\cite{pdg}.
The large value of $|V_{tb}|$ in the standard model implies that $R$,  
the ratio of branching fractions  
${B(t\rightarrow Wb)}/{B(t\rightarrow Wq)}$ 
(where $q$ is a $d$, $s$ or $b$ quark), is close to unity
and that the branching ratio for the decay of a top quark to $W b$ is 
nearly 100\%. 
This prediction has been used in the discovery of the top quark but
has not been, until now, confirmed experimentally.

In this letter we present the first direct measurement of $R$. The
result provides additional support for the top quark discovery and the 
first direct constraint on the CKM element $|V_{tb}|$ under the 
assumption of unitarity. The analysis is 
performed using 109 pb$^{-1}$ of 
proton-antiproton collisions data recorded at a center of mass energy of
1.8 TeV by the Collider Detector at Fermilab (CDF) during the
1992-1995 run of the Tevatron Collider at Fermilab. The CDF detector is
described in detail elsewhere~\cite{CDFNIM}; here we briefly 
describe only the
components which play a major role in this analysis.

The CDF tracking system consists of three different detectors embedded
in a 1.4~T solenoidal magnetic field.
A particle emerging from the interaction region passes 
through a four-layer
silicon vertex detector (SVX)~\cite{SVX,SVXI} located just outside 
the beam pipe, a set of vertex time projection chambers (VTX) 
and a drift chamber (CTC) with 84 measuring planes.
The CTC performs the pattern recognition and a three-dimensional 
reconstruction of charged particles. 
The VTX measures the position of the  primary interaction vertex 
along the beam axis.
Finally, the SVX, with its $r-\varphi$ readout in the plane 
perpendicular to the colliding beams, is designed
to determine precisely the impact parameter of the tracks.
A momentum resolution of 
$\Delta P_T/P^2_T \simeq  0.0009$ (GeV/c)$^{-1}$~\cite{PT} and an 
asymptotic impact parameter resolution of $ \simeq 13$ $\mu$m 
is obtained for
tracks with high $P_T$ detected by the SVX and the CTC.
Outside the tracking volume, electromagnetic and hadronic calorimeters
measure the energy of particles in the region $|\eta| \leq 4.2$~\cite{eta}.
Electron identification is obtained by combining calorimetric
and tracking information. Muon identification is performed by matching
tracks reconstructed in the CTC with segments measured by a system of drift 
chambers (muon chambers) located outside the calorimeters and 
covering the region $|\eta| < 1$.

The top quark has been observed~\cite{CDFI,CDFII,D0} only 
when produced in pairs. 
Assuming that the top quark decays to a real $W$ boson, it is customary
to classify $t \bar{t}$ final states according to the decay 
modes of the two W 
bosons.
We use two $t \bar{t}$ candidate data sets: the lepton+jets 
($l+jets$) and the dilepton samples.
The $l+jets$ sample, in which one $W$ decays to an electron or a muon and 
its corresponding neutrino and the  
other $W$ decays into two jets, has a final state characterized by
a high-$P_T$ lepton, missing transverse energy 
\mbox{\protect $\protect\not \! \! E_T$}~\cite{met}, and 
four jets.
The dilepton sample, in which both $W$ bosons decay into electrons or 
muons and neutrinos, is characterized by a final state with missing 
transverse energy, two high-$P_T$ leptons and two jets.
The selection criteria required to identify the $W$ bosons and to 
enhance the top quark content in these data sets
have been described in detail in previous 
CDF publications~\cite{CDFI,CDFII}.
To isolate the $l+jets$ sample we require the presence of one central 
($|\eta|<1$) lepton ($e$ or $\mu$) with $P_T> 20$ GeV/c, 
$\mbox{\protect $\protect\not \! \! E_T$} > 20$ 
GeV, three jets with $E_T > 15$~\cite{ET} GeV 
within $|\eta|< 2$ and a fourth 
jet with $E_T > 8$ GeV within $|\eta|< 2.4$. 
In the dilepton sample we require two leptons ($e$ or $\mu$) with 
$P_T>20$ GeV/c,
$\mbox{\protect $\protect\not \! \! E_T$} > 25$ GeV 
and two jets with $E_T >10$ GeV in the region $|\eta|< 2.0$. 
Candidate $Z$ events are removed by rejecting events containing same-flavor 
lepton pairs with  opposite charge whose invariant mass lies between 
75 and 105 $\textrm{GeV/c}^2$.
By construction the two data sets have no overlap.
After applying all the selection criteria we find 163 
events in the $l+jets$ and 9 events in the dilepton sample.

The presence of the $t \rightarrow W b$ decay is deduced by identifying 
jets associated with $b$ hadron decays using two distinct algorithms: the
SVX tagger and the SLT tagger.
The SVX tagging algorithm~\cite{CDFII} relies on the long lifetime of 
$b$ hadrons.
It searches for $b$ hadron decay vertices which are 
significantly displaced 
from the primary vertex and have three or more associated tracks.
If this search fails, tighter
quality cuts are applied to the tracks within the jet and vertices with
two tracks are also
accepted. In both cases, the transverse displacement of the decay 
vertex from the primary vertex, divided by its uncertainty, is 
required to be larger than 3.
The SVX algorithm is characterized by an
efficiency ($\varepsilon_{s}$) 
to tag a single $b$ jet in a $t \bar{t}$ event of 
($37.0 \pm 3.7$)\% and by a fake tagging rate of about 0.5\%. 
The SLT tagging is performed by looking for low-$P_T$ (relatively soft
compared to the primary lepton) muons and electrons
from semileptonic $b$ hadron decays.
The algorithm looks for low transverse momentum electron and muon
candidates
by matching CTC tracks with $P_T > 2$ $\textrm{GeV/c}$ 
with calorimeter clusters and 
track segments in the muon chambers. Moreover, 
to classify the event as a $t \bar{t}$ 
candidate, the soft lepton is required to be within $\Delta R < 
0.4$~\cite{deltar}
from one of the four highest-$E_T$ jets in the event.
The SLT algorithm has an efficiency per jet ($\varepsilon_{l}$) 
of ($10.2 \pm 1.0$)\% 
and a fake tagging rate of about 2\%.
Background due to fake tags is measured for both algorithms using samples
of QCD jet data sets~\cite{XSEC}.

The unknown ratio of branching fractions, $R$, is measured by 
comparing the 
observed number of tags in the data with expectations based on selection
criteria acceptances, tagging efficiencies and background estimates.
In the dilepton sample only SVX tagging is used and the sample is divided
into three non-overlapping bins: events with no $b$-tags 
(bin 0), one and only one $b$-tag (bin 1) and two $b$-tags (bin 2).
The use of SLT tagging in the dilepton data set does 
not provide any additional
statistical gain. 
In the $l+jets$ sample, we use both the SVX and SLT algorithms. Monte Carlo 
studies~\cite{MAS}
indicate that a superior use of the tagging information is obtained 
by dividing events into the same three bins used for the dilepton sample and 
then by subdividing 
the bin with no SVX tags into two bins according to the SLT tagging status.
The first bin (bin 00) is populated by events which are tagged 
by  neither the SVX nor the SLT algorithm and the second one (bin 01) 
contains events with one or more SLT tags and no SVX tags.

The number of observed events in each bin is reported 
in Table~\ref{tab:bins}.
The expected number of events, $N_i$, in each of 
the bins of the $l+jets$ sample can be expressed as a
function of the acceptances, tagging efficiencies and the 
estimated background, by the following set of equations:
\begin{mathletters}
\begin{eqnarray}
 N_{00} & = &  {n}_{0} + (1-\varepsilon_{l}) 
(1-\varepsilon_{s}) {n}_{1} +
     (1-\varepsilon_{l})^2 (1-\varepsilon_{s})^2 {n}_{2} + 
F_{00} \label{eq:nuno} \\
 N_{01} & = & \varepsilon_{l}  (1-\varepsilon_{s}) {n}_{1} +
  \varepsilon_{l} (2-\varepsilon_{l}) (1-\varepsilon_{s})^2 {n}_{2} 
+F_{01} \label{eq:nunob}
 \\
 N_{1\phantom{0}} & = & \varepsilon_{s} {n}_{1} + 2  
   \varepsilon_{s}  (1-\varepsilon_{s})  {n}_{2} + F_1  \label{eq:ndue}  \\
 N_{2\phantom{0}} & = & \varepsilon_{s} ^{2} {n}_{2} + F_2  \label{eq:ntre}
\end{eqnarray}
\end{mathletters}

\noindent with ${n}_i$ ($i=0,1,2$), the number of events with $i$ 
$b$-jets in the SVX acceptance, given by:
\begin{mathletters}
\begin{eqnarray}
 {n}_{0} & =  &
    N_{top}  [a_0 + (1- R) a_1 + (1-R)^2 a_2]   \label{eq:runo}\\
 {n}_{1} & = & N_{top}  [R a_1
   + 2   R  (1-R) a_2] \label{eq:rdue}\\
 {n}_{2} & = & N_{top}    R ^2 a_ 2\label{eq:rtre}
\end{eqnarray}
\end{mathletters}

\noindent where $N_{top}$ is the total number of $t \bar{t}$ events in the 
sample, $F_i$ is the background in the $i$-th bin and $a_i$ is the fraction 
of events containing $i$ $b$-jets ($i=0,1,2$) in the acceptance. 
This definition of acceptance, which reflects the way the $a_i$'s are
related to $R$ in Eq.~(\ref{eq:runo})--(\ref{eq:rtre}), 
has been chosen in
order to be able to use the standard CDF top Monte Carlo (see below) which
assumes $R=1$. 
For the dilepton
sample, Eq.~(\ref{eq:nuno}) and (\ref{eq:nunob}) are merged into one 
because SLT tagging is not used.

The unknown ratio $R$ is obtained by minimizing the negative 
logarithm of a likelihood function. Since the $l+jets$ and dilepton samples
are independent, the global likelihood can be written as:
\begin{equation}
 {\cal L} = {\cal L}_{\ell+jets}  {\cal L}_{\rm{dilepton}} \label{eq:lik}
\end{equation}

\noindent where each of the individual likelihoods is of the form:
\begin{displaymath}
 {\cal L}_{\alpha} = 
 \prod_{i} P(N_i;\bar{N}_i) \prod_{j} G(x_j;\bar{x}_j,\sigma_j)
\end{displaymath}

In this expression, $P(N_i;\bar{N}_i)$ is the 
Poisson probability for observing
$N_i$ events in each bin (the index $i$ runs from 1 to 4 for the $l+jets$
sample and from 1 to 3 for the dilepton one) 
with an expected mean $\bar{N}_i$ (see 
Table~\ref{tab:bins}). The functions $G(x_j;\bar{x}_j,\sigma_j)$ are Gaussians
in $x_j$, with mean $\bar{x}_j$ and variance $\sigma_j^2$, and 
incorporate the uncertainties in the tagging efficiencies, backgrounds and 
acceptances into the likelihood functions.

The acceptances and efficiencies are obtained using a $t \bar{t}$ Monte Carlo 
($M_{top}
=175$ $\textrm{GeV/c}^2$) data set generated using PYTHIA~\cite{PYT},
combined with a detailed simulation of the detector response.
The total number of $t \bar{t}$ pairs ($N_{top}$) in the two data samples is 
left as a free parameter.
The acceptances in each bin are normalized with respect to the bin with 
no $b$-jets. As a consequence, the trigger and 
lepton identification efficiencies cancel out in the ratio. 
We obtain $r_1 = 11.8\pm 1.2$ $(14.5 \pm 1.4)$ and 
$r_2 = 38.7\pm 3.9$ $(58.5 \pm 5.8)$, where 
$r_i = a_i/a_0$, for the $l+jets$ (dilepton) sample.
The uncertainties in these ratios include contributions from 
the jet energy scale and from the Monte Carlo modeling of 
initial and final state radiation.

The background in the untagged sample is mainly due to the associated 
production of $W$ bosons with light quark jets.
The backgrounds to the SLT and SVX tagged events (background in
bin 01 and 1, respectively), are mainly due to the associated 
production of $W$ bosons and heavy quarks ($W b \bar{b}$, 
$W c \bar{c}$, $Wc$) 
and to mistags due to mismeasured tracks. Smaller contributions 
come from $b \bar{b}$,
diboson production ($WW$, $ZZ$ and $WZ$), $Z \rightarrow \tau \tau$ decays, 
Drell-Yan lepton pair production and single top quark production.
These backgrounds are calculated using a combination of data and 
Monte Carlo information~\cite{XSEC,MAS}.
The initial values of the SVX and SLT backgrounds are a function of the 
$t \bar{t}$ content of the $l+jets$ sample itself, and therefore 
need to be appropriately corrected~\cite{CDFI}.
An iterative process is used to account for this effect and has been
implemented in the likelihood minimization procedure used to
estimate $R$. Using this procedure, as output of the likelihood 
minimization, we 
estimate $F_{1}=3.3^{+2.3}_{-1.2}$ 
and  $F_{01} = 7.2 \pm 1.6$ events for the SVX and SLT backgrounds, 
respectively. In the same way, the
background to double SVX tagged events (bin 2) is estimated to be small 
and amounts to $F_{2} = 0.2 \pm 0.1$ events. 
The background in bin 00, $F_{00}$, is obtained as follows.
Defining $N_{tot}$ to be the total number of events in the $l+jets$
data set, $N_{SVX}$ the total number of SVX tagged events in this
sample, $F_{SVX}$ the estimated background and $\epsilon_{SVX}$ the
SVX event tagging efficiency, the total number of top events is:
$N_{t \bar{t}}=(N_{SVX}-F_{SVX})/(\epsilon_{SVX} R)$. The number of 
background events before tagging 
is given by: $F = N_{tot}-N_{t \bar{t}}$ and
therefore the background in bin 00 is $F_{00}=F-(F_{01}+F_{1}+F_{2})$. 
As before, the estimate is performed iteratively during the likelihood
minimization. 

The initial background to the dilepton sample has been estimated to be
$2.4\pm0.5$~\cite{dilept}. 
In this sample, we estimate a
background of $0.10\pm0.04$ events to SVX single tagged events. 
The double SVX tagged background is negligible 
($F_2=0$ in Eq.~(\ref{eq:ntre})).
In this case,
the number of background events is not a function of the $t \bar{t}$ content
of the initial sample and no special correction need to be applied.
As in the $l+jets$ case, the background in bin 0 is obtained by a subtraction 
of the tagged background from the total background and amounts to $2.3\pm0.5$ 
events. 
The resulting number of background events after the likelihood minimization
procedure is shown in Table~\ref{tab:back} for both data sets.

\begin{figure}
\centering
\epsfig{figure=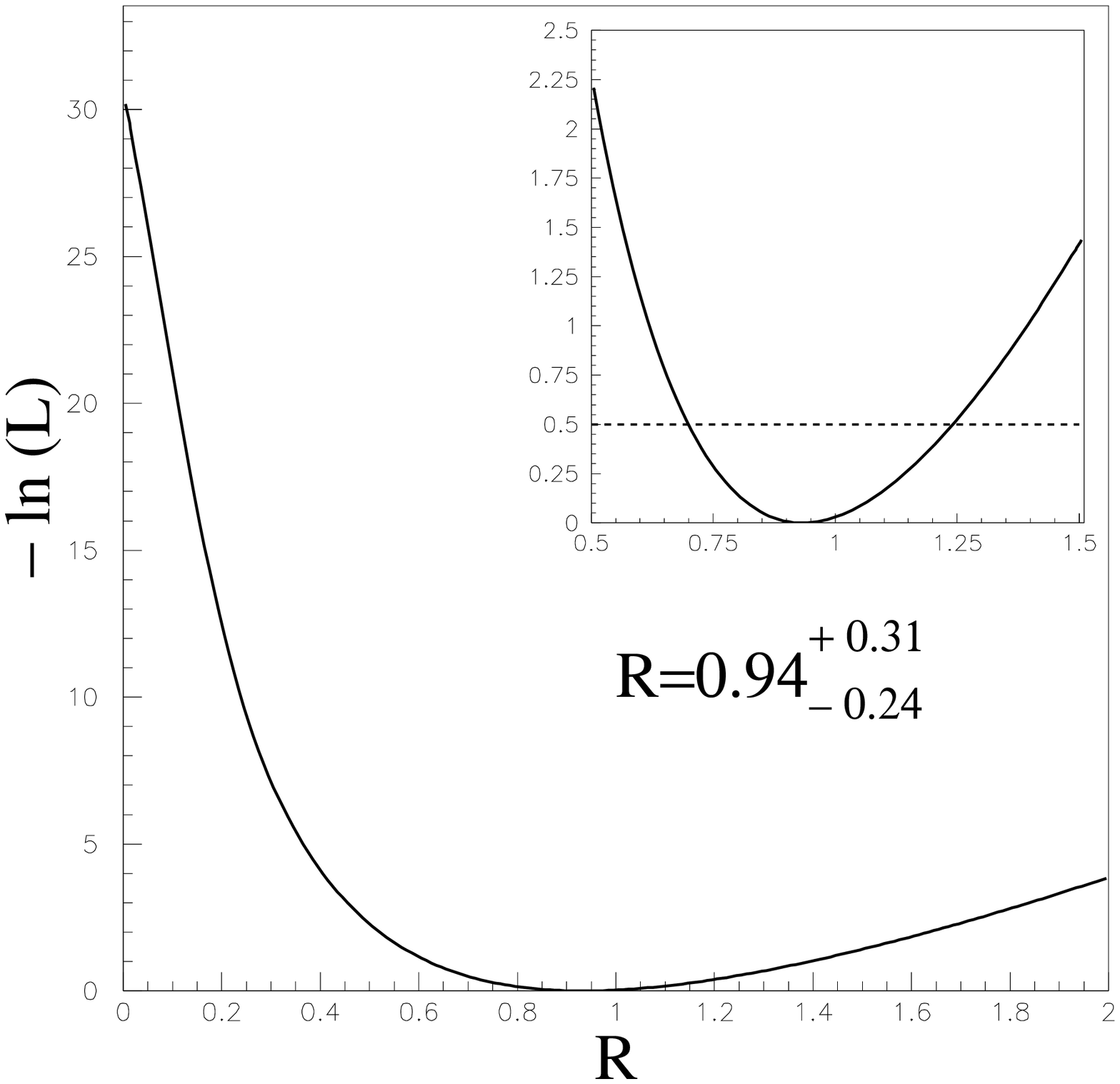,width=\linewidth}
\caption{The negative logarithm of the likelihood function
of Eq.\ (\ref{eq:lik}) as a function of R. The inset plot is a 
magnified view of the region around the minimum.}
\label{fig:like} 
\end{figure}

The likelihood minimization yields $R ={0.94}^{+0.31}_{-0.24}$. 
The uncertainties 
includes both statistical and systematic effects with the former
being the dominant contribution.
The negative log-likelihood as a function of $R$ is shown in 
Fig.\ \ref{fig:like}. 
The lower limit on $R$ is obtained by a numerical integration of the 
likelihood function
and we obtain: $R > 0.61$ $(0.56)$ at 90 (95)\% CL.
 
The CKM element $|V_{tb}|$ is directly related to $R$, although in a 
model-dependent way. We
assume that the top quark decays to non-$W$ final states are 
negligible~\cite{FCNC,Higgs}.
Under this assumption $R$ is related to $|V_{tb}|$ by:
\begin{equation}
R = \frac{|V_{tb}|^2}{|V_{ts}|^2+|V_{td}|^2+|V_{tb}|^2}. \label{eq:b2vtb}
\end{equation}

\noindent If we assume three 
generation unitarity, the denominator is equal to unity and therefore
$R = |V_{tb}|^2$. As a consequence, we obtain:
$|V_{tb}| = {0.97}^{+0.16}_{-0.12}$ or $|V_{tb}| > 0.78$ $(0.75)$ 
at 90 (95)\% CL.

The result, although limited by statistics, represents the first direct 
measurement of $R$. The large value of $R$ that we measure
is consistent with standard model expectations and supports the 
assumption that top quarks decay predominantly to $b$ quarks.
Under the assumption of three generation unitarity, our
calculated value of 
$|V_{tb}|= {0.97}^{+0.16}_{-0.12}$ ($|V_{tb}| > 0.78$ at 90\% CL) 
is consistent
with indirect limits obtained from global fits.

We thank the Fermilab staff and the technical staffs of the participating
institutions for their vital contributions. This work was supported by the
U. S. Department of Energy and National Science Foundation, the Italian
Istituto Nazionale di Fisica Nucleare, the Ministry of Education, Science
and Culture of Japan, the Natural Sciences and Engineering Research Council
of Canada, the National Science Council of the Republic of China, and the 
A. P. Sloan Foundation.

\begin{table}[htb]
\centering
\begin{tabular}{lcccc}
sample & 00 & 01 & 1 & 2 \\
\hline
$l+jets$    & 126 & 14   &  18  & 5 \\
dilepton    & 6   & n/a  &  3   & 0 \\ 
\end{tabular} 
\caption{Number of events for the two data samples in each of the bins 
defined according to the number of SVX and SLT tags.
The case of SLT tags (bin 01=``one or more SLT tag, no SVX tag'') 
does not apply (``n/a'') to the dilepton data set (see text).}
\label{tab:bins}
\end{table}

\begin{table}[htb]
\centering
\begin{tabular}{lcccc}
sample & 00 & 01 & 1 & 2 \\
\hline
$l+jets$    & $108\pm10$ & $7.2\pm1.6$  &$3.3^{+2.3}_{-1.2}$ & 
   $0.2\pm0.1$ \\
dilepton    & $2.3\pm0.5$   & n/a &  $0.10\pm0.04$ & n/a \\ 
\end{tabular} 
\caption{Estimated number of background events bin by bin for the two 
data samples. The case of SLT tags (bin 01) applies only to the 
$l+jets$ data set and the double tagged dilepton background is 
neglected (``n/a'').}
\label{tab:back}
\end{table}


\begin{references}
\bibitem{PT} The transverse momentum, $P_T$, is defined as the projection of
the momentum onto the plane transverse to the beam axis.
\bibitem{eta} The pseudorapidity, $\eta$, is defined as
$\eta = -\ln[\tan(\theta/2)]$, where $\theta$ is the polar angle with 
respect to the proton beam direction ($z$-axis).
\bibitem{met} The missing transverse energy, 
\mbox{\protect $\protect\not \! \! E_T$}, 
is defined as the
negative of the vector sum of transverse energy in all calorimeters
towers with $|\eta|<3.6$.
\bibitem{ET} The transverse energy, $E_T$, is defined as the projection of
the energy onto the plane transverse to the beam axis.
\bibitem{deltar} We define 
$\Delta R \equiv \sqrt{(\Delta \phi)^2+(\Delta \eta)^2}$,
where $\Delta \eta$ and $\Delta \phi$ refer to the difference
in pseudorapidities and azimuthal angles respectively of the two
objects in question.
\end{references}
\end{document}